\documentclass[showpacs,amsmath,amssymb,aps,pra,10pt,reprint,superscriptaddress]{revtex4-1}
\usepackage[breaklinks=true,colorlinks=true,linkcolor=blue,urlcolor=blue,citecolor=blue]{hyperref}
\usepackage{graphicx}
\usepackage{epstopdf}
\usepackage{soul,xcolor}
\usepackage{mathptmx} 

\begin{document}
\draft

\title{AC losses in macroscopic thin-walled superconducting cylinders}

\author{M.\,I. Tsindlekht}
    \email[Corresponding author: ]{mtsindl@mail.huji.ac.il}
\author{V.\,M. Genkin}
\author{I. Felner}
\author{F. Zeides}
\author{N. Katz}
    \affiliation{The Racah Institute of Physics, The Hebrew University of Jerusalem, 91904 Jerusalem, Israel}
\author{$\check{\text{S}}$.~Gazi}
\author{$\check{\text{S}}$. Chromik}
    \affiliation{The Institute of Electrical Engineering SAS, D$\acute{u}$bravsk$\acute{a}$ cesta 9, 84104  Bratislava, Slovakia}
\author{O.\,V. Dobrovolskiy}
    \affiliation{Cryogenic Quantum Electronics, Institut f\"ur Elektrische Messtechnik und Grundlagen der Elektrotechnik, Technische Universit\"at Braunschweig, Hans-Sommer-Str. 66, 38106 Braunschweig, Germany}
    \affiliation{Laboratory for Emerging Nanometrology, Technische Universit\"at Braunschweig, Langer Kamp 6a/b, 38106 Braunschweig, Germany}
    \affiliation{University of Vienna, Faculty of Physics, Nanomagnetism and Magnonics, Superconductivity and Spintronics Laboratory, W\"ahringer Str. 17, 1090 Vienna, Austria}
\date{\today}

\begin{abstract}
Measurements of the ac response represent a valuable method for probing the properties of superconductors. In the surface superconducting state (SSS), a current exceeding the surface critical current $I_\mathrm{c}$ leads to breakdown of SSS and penetration of external magnetic field into the sample bulk. An interesting free-of-bulk system in SSS is offered by thin-walled cylinders. According to the full penetration of magnetic flux (FPMF) model, each time the instant value of an ac field is equal to a certain critical value, the ac susceptibility $\chi$ will exhibit \emph{jumps} as a function of the ac field amplitude $H_\mathrm{ac}$ because of the periodic destruction and restoration of SSS in the cylinder wall. Here we study the low-frequency (128-8192\,Hz) ac response of thin-walled niobium cylinders under superimposed dc and ac magnetic fields applied parallel to the cylinder axis. In contrast to the FPMF model predictions, experiments reveal a \emph{smooth} $\chi(H_\mathrm{ac})$ dependence. To explain the experimental observations, we propose a phenomenological \emph{partial} penetration magnetic flux (PPMF) model, which assumes that after restoration of the superconducting state, the magnetic fields inside and outside the cylinder are unequal and the magnitude of the penetrating flux is random for every penetration.
This model fits very well the experimental data on the temperature dependence of the first harmonic $\chi_1$ for any dc field and ac amplitude.
\end{abstract}
\date{\today}
\maketitle

\section{Introduction}
The ac magnetic response of superconductors contains essential information on their properties and it has been a matter of extensive research for both conventional and high-$T_\mathrm{c}$ superconductors\,\cite{Max63prl,Rol67prv,Hop74prb,Ish90prb,Gom97sst,You09jap}. Studies of low-$T_\mathrm{c}$ superconductors revealed an absorption maximum near $T_\mathrm{c}$ upon variation of temperature and dc magnetic field\,\cite{Max63prl,Rol67prv,Hop74prb}. The ac response of high-$T_\mathrm{c}$ superconductors exhibits similar features\,\cite{Ish90prb,Gom97sst,You09jap}. However, the physical reasons for this behavior remain unclear so far.

The first Maxwell-Strongin model\,\cite{Max63prl} assumed the existence of microscopic superconducting filaments that causes the increase of normal conductivity\,\cite{Doi64phl}. A more general eddy current model\,\cite{Cod68phr} suggested that the normal conductivity increases due to the appearance of superconducting inclusions, which leads to the absorption maximum near $T_\mathrm{c}$. However, an analysis based on the two-fluid model did not confirm the existence of an absorption maximum in the ac response\,\cite{Kho83pla}.
Analysis of the ac response of high $T_\mathrm{c}$ single crystals with weak volume pinning in ac and dc magnetic fields normal to the sample plane showed that the ac susceptibility at the first and especially the third harmonics is determined by the Bean-Livingston\,\cite{Bea64prl} and geometric barriers\,\cite{Zel94prl}, but not by bulk pinning\,\cite{Bee96pcs}. On the other hand, the BCS theory\,\cite{Tin04boo} asserts an increase in the normal conductivity at low frequencies due to a singularity in the density of states\,\cite{Kho83pla}. This dissipative conductivity has a logarithmic singularity and only appears in a vanishingly narrow temperature interval\,\cite{Kho83pla}.

In magnetic fields whose magnitudes are between the second and third critical fields, $H_\mathrm{c2}< H_0 <H_\mathrm{c3}$, a thin surface sheath remains superconducting\,\cite{Gen66boo}, giving rise to the surface superconducting state. In the SSS, ac losses were observed in superimposed dc and ac external fields $H_\mathrm{ext}(t) \equiv H_0+H_\mathrm{1}(t)= H_0 + H_\mathrm{ac}\sin(\omega t)$ applied parallel to the sample surface at dc magnetic field $H_0> H_\mathrm{c2}$\,\cite{Rol67prv,Str64prl,Hop74prb,Tsi11prb}. Here, $H_\mathrm{ac}$ and $\omega$ are the amplitude and frequency of the ac magnetic field and $t$ is the time.

\begin{figure}[t!]
    \centering
    \includegraphics[width=1\linewidth]{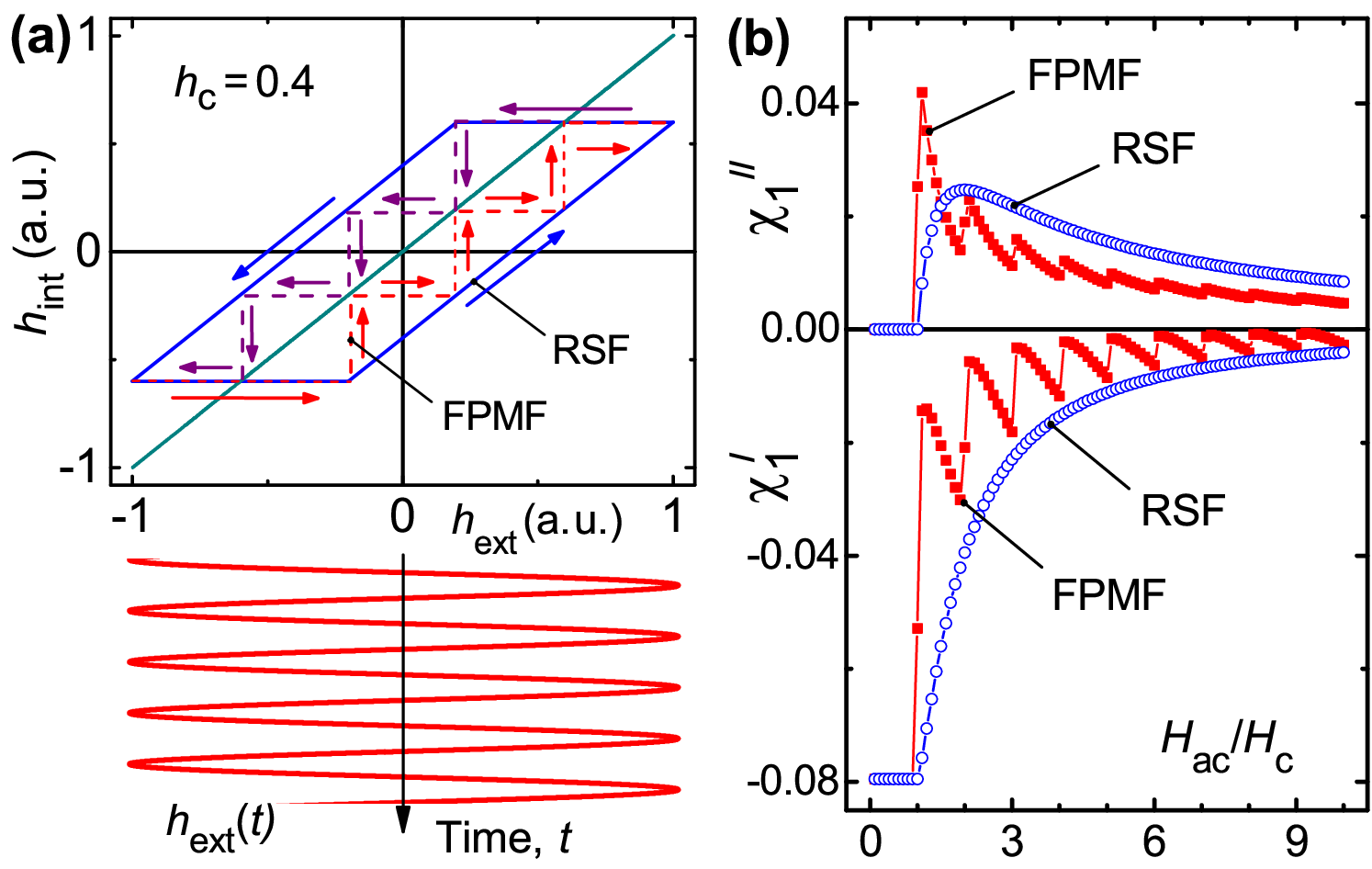}
    \caption{Qualitative comparison of the FPMF and RSF model predictions. (a) Minor magnetization loops $h_\mathrm{int}(h_\mathrm{ext}(t))$ of a superconductor during a slow course of the ac cycle and
    (b) field amplitude dependence of the ac susceptibility $\chi_1 = \chi_1^{\prime} + i \chi_1^{\prime\prime}$.}
    \label{f1}
\end{figure}

It is known that in the presence of a transport current $I$, the SSS is stable only if $I<I_\mathrm{c}$, where $I_\mathrm{c}$ is the surface critical current\,\cite{Abr65etp,Fin65prl,Par65prl}. If $I$ reaches $I_\mathrm{c}$ at a certain instant value of $H_\mathrm{ext}(t)$, a further increase of the field leads to destruction of the SSS. In this case, full penetration of the external field into the sample bulk is expected\,\cite{Rol67prv}, which implies that the internal field $H_\mathrm{int}(t)$ becomes equal to $H_\mathrm{ext}(t)$. Within the framework of the \emph{full penetration of magnetic flux} (FPMF) model, upon a slow variation of $H_\mathrm{1}(t)$, a minor magnetization loop $h_\mathrm{int}(h_\mathrm{ext}(t))$ should feature \emph{steps}, as shown by the dashed line in Fig.\,\ref{f1}(a). Here and in what follows we designate the internal field $H_\mathrm{int}(t)\equiv H_\mathrm{int 0} +H_\mathrm{int1}(t)$ and use the dimensionless quantities $h_\mathrm{int}(t) = H_\mathrm{int1}/H_\mathrm{ac}$, $h_\mathrm{ext}(t)=H_1(t)/H_\mathrm{ac}$ and $h_\mathrm{c}=H_\mathrm{c}/H_\mathrm{ac}$, here $H_\mathrm{c}$ is maximum difference, $max(H_\mathrm{ext}(t) - H_\mathrm{int}(t))$, between the values of the external and internal magnetic fields. If this difference is greater than $H_c$, the superconducting state will be destroyed. This maximum difference corresponds to the maximum surface current. Accordingly, for such a stepped minor magnetization loop the ac susceptibility $\chi_1(H_\mathrm{ac})$ should exhibit \emph{jumps}, see Fig.\,\ref{f1}(b). However, such instabilities in $\chi_1(H_\mathrm{ac})$ and $H_\mathrm{int}(H_\mathrm{ext}(t))$ have not yet been observed experimentally, see, for example,\,\cite{Rol67prv}.

By contrast, the \emph{Rollins-Silcox-Fink} (RSF) model\,\cite{Rol67prv,Fin66prl} based on a \emph{parallelogram-shaped} minor magnetization loop $H_\mathrm{int}(H_\mathrm{ext}(t))$ predicts a \emph{smooth} $\chi_1(H_\mathrm{ac})$ curve, see Fig.\,\ref{f1}(b). The RSF model assumes that at $H_0>H_\mathrm{c2}$, as soon as the difference between $H_\mathrm{ext}(t)$ and $H_\mathrm{int}(t)$ reaches a critical value $H_\mathrm{c}$, the current in the entire surface sheath becomes equal to $I_\mathrm{c}$. In this case $H_\mathrm{int}(t)$ follows $H_\mathrm{ext}$ with some delay, i.e. $H_\mathrm{int}=H_\mathrm{ext}\mp H_\mathrm{c}$. The signs of $H_\mathrm{c}$ correspond to increasing and decreasing external field, respectively. This assumption implies that for $|H_\mathrm{ext} - H_\mathrm{int}| < H_\mathrm{c}$ the surface current is a linear function of $H_\mathrm{ext}$ and $H_\mathrm{int}$ does not depend on $H_\mathrm{ext}$\,\cite{Rol67prv,Fin69prv,Fin67prv}. Accordingly, during a slow course of the ac cycle at a constant $H_\mathrm{c}$, a minor magnetization loop $H_\mathrm{int}(H_\mathrm{ext}(t))$ is parallelogram-shaped. However, the RSF model does not account for the destruction and restoration of superconductivity in the SSS.

A hollow thin-walled superconducting cylinder in the presence of $H_\mathrm{ext}$ applied parallel to the cylinder axis is a model for a bulk sample in the SSS. At the same time, this object is free of complications associated with the presence of the normal state in the sample bulk. Namely, if the external field increases slowly and the field in the cylinder hollow does not change, the current in the wall increases and can reach the critical value $I_\mathrm{c}$. A further increase of $H_\mathrm{ext}$ leads the wall to transit to the normal state, and the order parameter is equal to zero. The external magnetic field penetrates then into the hollow and the current in the wall decreases to zero. The normal state of the wall at $I=0$ is unstable since $T<T_\mathrm{c}$ and superconductivity in the wall is restored.

The answer to the question of whether surface superconducting states exist in type II thin films was obtained almost 60 years ago\,\cite{Bur65prv}. In particular, the authors showed that screening of an ac field in a thin-walled cylinder begins in a magnetic field below $H_\mathrm{c3}$. In our previous work\,\cite{Tsi14prb} and in the present study we demonstrate that screening in a field parallel to the surface of a thin-walled cylinder exists in dc magnetic fields larger than $H_\mathrm{c2}$. Here, $H_\mathrm{c2}$ is deduced from the magnetization curve $M(H)$ and, with the accuracy of our experiment, $M$ is equal to zero at $H > H_\mathrm{c2}$. The magnetic-field dependence of the ac susceptibility shows that ac losses appear in magnetic fields $H > H_\mathrm{c2}$. Apparently, at $H<H_\mathrm{c2}$ the vortices are pinned. The thickness of the studied samples, $d = 120$\,nm is quite large so that the superconducting coherence length $\xi < d$ at low temperatures and $\xi \geq d$ only near $T_\mathrm{c}$. It should be noted that at a temperature when $\xi = d$, $H_\mathrm{c2}$ is equal to 220\,Oe. This occurs very close to $T_\mathrm{c}$.

Here, we present an experimental study of the low-frequency (128-8192\,Hz) ac magnetic response of macroscopic thin-walled superconducting cylinders under superimposed dc and ac magnetic fields applied parallel to the cylinder axis. In contrast to the predictions of the FPMF model, our experiments reveal that the ac susceptibility $\chi_1$ of thin-walled cylinders is a \emph{smooth} function of $H_\mathrm{ac}$. To explain this, we propose a phenomenological \emph{partial penetration of magnetic flux} (PPMF) model which implies that after a restoration of the superconducting state in the wall, the magnetic fields inside and outside the cylinder are \emph{not equal}, and the value of the penetrating flux is random for each penetration. In this case $\chi_1$ is a smooth function of $H_\mathrm{ac}$. The proposed PPMF model is in very good agreement with the experimental data for the first-harmonic ac susceptibility in the entire temperature range under any dc field magnitudes and ac excitation amplitudes. However, in a certain temperature range, the deduced parameters using the PPMF model are questionable.

\section{Samples and setup}
The samples are two thin-walled superconducting cylinders with rectangular cross-sections. The samples were prepared by dc magnetron sputtering of niobium on rotating sapphire substrates with rounded corners (Gavish Sapphire Products, Israel). The dimensions of the samples are $1.4\times 3 \times 12$\,mm$^3$ (sample F15) and $1.4\times 3 \times 21$\,mm$^3$ (sample GL). The wall thicknesses are $300$\,nm and $120$\,nm, respectively.

\begin{figure}[t!]
    \centering
    \includegraphics[width=0.95\linewidth]{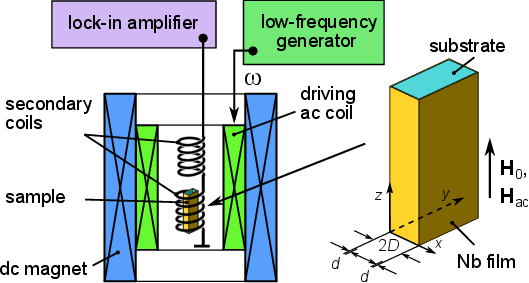}
    \caption{Block-diagram of the experimental setup and the sample geometry. Both dc and ac magnetic fields are applied parallel to the cylinder axis.}
    \label{f2}
\end{figure}

The magnetic moment oscillations of the samples were measured at the first, second and third harmonics of the driving ac frequency using the pickup coils method\,\cite{Lev05prb}. The dc and ac magnetic fields were applied parallel to the axes of the cylinders, see Fig.\,\ref{f2} for the geometry. The amplitude and phase of the imbalanced signal were measured with a lock-in amplifier. A homemade measurement cell was adapted to a commercial SQUID magnetometer. The block diagram of the setup is also shown in Fig.\,\ref{f2}. The temperature dependences of the signals at the first, second and third harmonics of the excitation frequencies were measured concurrently in two modes. (i) In the temperature-sweep mode, the temperature was swept at a rate of 0.1\,K/min, with a continuous measurement of the magnetic response. The sweep rate 0.1\,K/min was low enough to avoid a lag between the actual sample temperature and the temperature measured by the sensor, as confirmed experimentally. (ii) In the second mode, temperature was changed point-by-point, with temperature stabilization at each point. Measurements in both modes at any dc magnetic field were carried out as the temperature decreased from $T>T_ \mathrm{c}$, i.e. in field cooling regime. The amplitude $H_\mathrm{ac}$ of the ac magnetic field was 0.005-0.2\,Oe. The second-harmonic signal was negligible, and we will not discuss its behavior here.

\begin{figure*}[t!]
    \centering
    \includegraphics[width=0.88\linewidth]{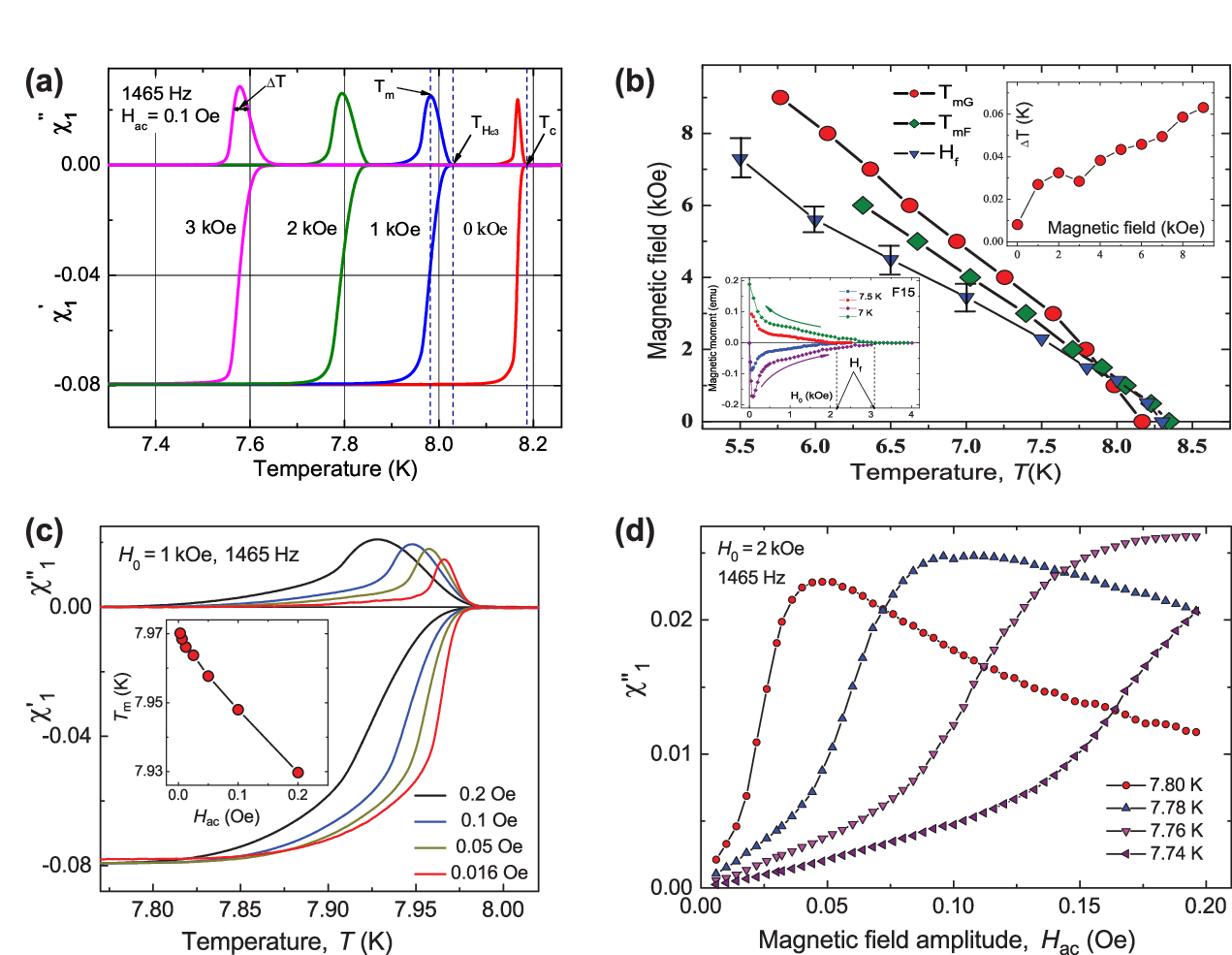}
    \caption{
    (a) $\chi^{\prime}(T)$ and $\chi^{\prime\prime}(T)$ for a series of dc field values for sample GL.
        $T_\mathrm{m}$: temperature at the maximum of absorption; $T_\mathrm{H_\mathrm{c3}}$: transition temperature at which the superconducting state appears in a finite dc magnetic field; $\Delta T$: linewidth of the absorption peak.
    (b) Dependence of $T_\mathrm{m}$ on the magnetic field and $H_\mathrm{f}$ on the temperature in the $H$-$T$ plane.
        $T_\mathrm{mF}$ and $T_\mathrm{mG}$ are the $T_\mathrm{m}$ values for samples F15 and GL, respectively.
        Lower inset: Ascending and descending branches of the F15 sample magnetization curve at T=7 and 7.5 K after zero field cooling. Upper inset: Field dependence of the linewidth $\Delta T$ for sample GL.
    (c) $\chi_1^{\prime}(T)$ and $\chi_1^{\prime\prime}(T)$ for a series of excitation amplitudes for sample GL.
        Inset: $T_\mathrm{m}$ versus $H_\mathrm{ac}$ at $H_0 = 1$\,kOe.
    (d) $\chi_1^{\prime\prime}(H_\mathrm{ac})$ for a series of temperatures near $T = 7.76$\,K for sample GL.
        The measurements were carried out in the swept temperature (a-c) and the point-by-point (d) modes.}
    \label{f3}
\end{figure*}

The ac magnetic moment of a sample in the ac magnetic field $H_1 =H_\mathrm{ac}\exp(-i\omega t)$ is given by
\begin{equation}
    \label{Eq1}
    M=VH_\mathrm{ac}\sum \chi_n\exp(-in\omega t),
\end{equation}
where $V$ is the sample volume and $\chi_n=\chi_n(\omega,H_\mathrm{ac})$ is the ac magnetic susceptibility at the $n$-th harmonic of the driving ac frequency $\omega$. The ac field is shielded completely at low temperatures. This allows the deduction of the absolute value of the in-phase and out-of-phase components of the ac magnetic susceptibility at all temperatures and magnetic fields.

\section{Experiment}
The experimentally measured dependences of the magnetic susceptibility as a function of temperature $T$, dc magnetic field magnitude $H_0$ and ac magnetic field amplitude $H_\mathrm{ac}$ are presented in Fig.\,\ref{f3}. Here, $\chi_1^{\prime}$ and $\chi_1^{\prime\prime}$ are real and imaginary parts of the complex ac susceptibility $\chi_1=\chi_1^{\prime} + i\chi_1^{\prime\prime}$ measured at the first harmonics of the excitation frequency. The dependences $\chi_1^{\prime}(T)$ and $\chi_1^{\prime\prime}(T)$ for sample GL are shown in Fig.\,\ref{f3}(a) for a series of dc magnetic field magnitudes. It can be seen that losses appear with decreasing temperature at $T= T_\mathrm{H_\mathrm{c3}}$. At this temperature $H_0=H_{c3}$, and at $H_0=0$ this temperature is equal to $T_c$.  With increase of $H_0$, the absorption peak, with a maximum at the temperature $T_\mathrm{m}$, shifts towards lower temperatures and its linewidth $\Delta T$ increases, see the upper inset in Fig.\,\ref{f3}(b). Here, the linewidth $\Delta T$ was deduced at $70\%$
of the maximal value of $\chi_1^{\prime\prime}$. We note that the linewidth is $\Delta T < 0.1$\,K in the investigated range of magnetic fields. This allows us to deduce the third critical field $H_\mathrm{c3} = H_0|_{T = T_\mathrm{m}}$ with high accuracy because $\Delta T\ll T_\mathrm{m}$, and the changes of $T_\mathrm{m}$ as a function of $H_\mathrm{ac}$ are very small, see the inset in Fig.\,\ref{f3}(c). We also note that $T_\mathrm{m}$ is practically independent of frequency in the investigated range 128-8192\,Hz (not shown).

\begin{figure}[t!]
    \centering
    \includegraphics[width=0.85\linewidth]{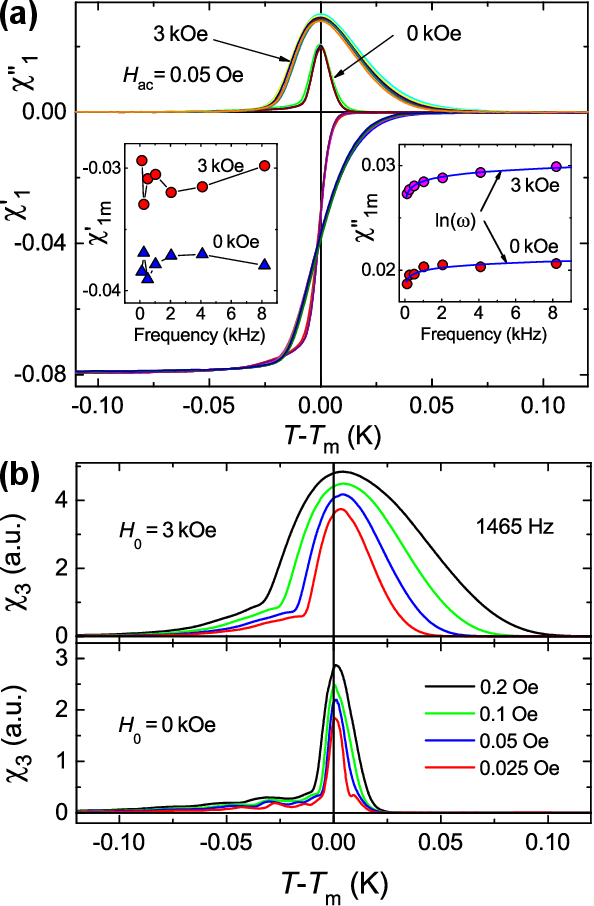}
    \caption{(a) $\chi_{1}^{\prime}(T)$ and $\chi_1^{\prime\prime}(T)$ at $H_0=0$ and 3\,kOe for seven excitation frequencies in the range $128-8192$\,Hz for sample GL.
    Insets: Frequency dependences of $\chi_1^{\prime}$ and $\chi_{1}^{\prime\prime}$ at $T = T_\mathrm{m}$ for $H_0=0$ and 3\,kOe.
             (b) $\chi_{3}(T)$ and $\chi_3(T)$ at $H_0=0$ and 3\,kOe for a series of ac amplitudes for sample GL.}
    \label{f4}
\end{figure}

The total magnetic moment of a cylinder in a parallel dc magnetic field is determined by the difference between the magnetic field in the hollow volume and the applied external field, since the superconducting thin wall volume is very small. The lower inset in Fig.\,\ref{f3}(b) shows an ascending branch of the magnetization curve for sample F15, from where we deduce the dc magnetic field $H_\mathrm{f}$ at which the dc magnetic moment becomes equal to zero. Similar measurements of magnetic moment of sample GL are impossible in our magnetometer for technical reasons. It should be noted that at $H_0 = H_\mathrm{f}$ the internal dc magnetic field becomes equal to the external field. From Fig.\,\ref{f3}(b) it is clear that $T_\mathrm{m}>T(H_\mathrm{f})$ .
Not so far from $T_c$ $H_\mathrm{f}$ varies as $1-T/T_\mathrm{c}$. Assuming that $H_\mathrm{f} = H_\mathrm{c2}$, then we can estimate the value of the coherence length $\xi=(\phi_0/2\pi H_{c2})^{1/2}$, here $\phi_0$ is the flux quantum. It turns out that $\xi\approx 32$ and 38~nm at $T =$ 7.0 and 7.5 K, respectively. At $T/T_c=0.998$  the coherence length becomes equal to the film thickness for sample F15, $\xi(T/T_c\approx 0.998)\approx~300~nm$. It is clear that $\xi(T_m)$ becomes less than the film thickness $d$ for $H_0>50$ Oe, assuming that $\xi(T) \propto 1/\sqrt{(T_c-T) }$. 

Figure\,\ref{f3}(c) shows how the excitation amplitude $H_\mathrm{ac}$ affects $\chi_1^{\prime}(T)$ and $\chi_1^{\prime\prime}(T)$. An increase of $H_\mathrm{ac}$ leads to a broadening and a shift of the absorption peak towards lower temperatures. The amplitude dependence of $T_\mathrm{m}(H_\mathrm{ac})$ is shown in the inset in Fig.\,\ref{f3}(c). The amplitude dependence of $\chi_1^{\prime\prime}$ in Fig.\,\ref{f3}(d) at a constant temperature shows that with the resolution of our experiment there are no jumps corresponding to the full penetration of the magnetic flux, see Fig.\ ,\ref{f1}(b). And besides, it shows that even a very small change in temperature, $0.02 K$, leads to large changes in $\chi_1^{\prime\prime}(H_\mathrm{ac})$.

In Fig.\,\ref{f4}(a) we show the real and imaginary parts of $\chi_1$ as a function of temperature for sample GL for seven frequencies from 128 to 8192\,Hz and the ac amplitude 0.05\,Oe in zero and 3\,kOe dc magnetic fields. The respective frequency dependences of $\chi_1^{\prime}$ and $\chi_1^{\prime\prime}$ at $T = T_\mathrm{m}$ are shown in the insets in Fig.\,\ref{f4}(a). The frequency dispersion of $\chi_{1}^{\prime}$ is small and  $\chi_{1}^{\prime\prime}(\omega)~\propto\ln(\omega)$. The frequency dispersion of the ac response requires a careful investigation which is beyond the scope of the present work.

Figure\,\ref{f4}(b) presents the magnitude of the susceptibility at the third harmonics of the ac driving frequency, $\chi_3(T)$, for a series of ac amplitudes at two dc magnetic field values. We do not observe zeroing of the third-harmonic signal with decreasing ac amplitude. An analysis shows that the perturbation theory is not applicable to the third-harmonic signal. It should be noted that a linear ac response near the normal state transition is unattainable in bulk samples too, even then the excitation amplitude is less than $1 mOe$, as was demonstrated in Ref.\,\cite{Tsi08prb}.

\section{Theoretical models}

In this section, we will dwell more on the theoretical models which may be considered for treating the experimental data. In particular, we will start with rough estimations for the effective penetration length and the skin depth within the two-fluid model. These estimations will, however, reveal that the existence of normal electrons cannot explain the experimental observations. Then, we will proceed to an analysis of the stability of the superconducting phase within the Ginzburg-Landau approach. It will be shown that the Ginzburg-Landau approach predicts a transition to the normal state to occur at a very small instantaneous value of the ac field, which is much less than the amplitude of the ac field in our experiments. Accordingly, to resolve inconsistencies between the available models, a new phenomenological PPMF model will be introduced and successfully used for fitting of the experimental data. In essence, the PPMF model can be viewed as a compromise between the FPMF and RSF models, providing a link between them.

\subsection{Two-fluid model}
As a model we consider an infinite dielectric slab of thickness $2D$ wrapped with superconducting films of thickness $d$ on both surfaces, thus $D\gg d$, see Fig.\,\ref{f2}, and the dc and ac magnetic fields applied along the $z$-axis. Within the framework of the two-fluid model and neglecting any nonlinear effects, for the current density $j(\omega)$ and the vector potential $A(\omega)$ in the superconducting film we use the following equations\,\cite{Tin04boo}
\begin{equation}
    \begin{array}{l}
    \label{Eq2}
    j(\omega)=-\displaystyle\frac{c}{4\pi}\left[\frac{1}{\lambda^2(\omega)}-\frac{2i}{\delta^2(\omega)}\right]A(\omega)\equiv -\frac{c}{4\pi\tilde\lambda^2}A(\omega),\\
    \\
    \displaystyle\frac{d^2A(\omega)}{dx^2} = \displaystyle-\frac{4\pi}{c}j(\omega).
    \end{array}
\end{equation}
In Eq.\,(\ref{Eq2}), the parameters $\lambda(\omega)$ and $\delta(\omega)$ characterize the \textit{effective} penetration length and the skin depth of the magnetic field in the film due to the superconducting and normal components of the conductivity. The axis $x$ is chosen perpendicular to the slab surface, as shown in Fig.\,\ref{f2}. The solution of Eq.\,(\ref{Eq2}) with gauge $A|_{x=0}=0$ and the boundary conditions for the continuity of the magnetic field on both sides of the films yields the magnetic field inside the slab
\begin{equation}
    \label{Eq3}
    \displaystyle\frac{H_\mathrm{int1}}{H_\mathrm{ac}} \equiv 1+4\pi\chi_1= \displaystyle\frac{1}{[\cosh(d/\tilde\lambda)+(D/\tilde\lambda)\sinh(d/\tilde\lambda)]}.
\end{equation}

\begin{figure}[t!]
    \centering
    \includegraphics[width=0.95\linewidth]{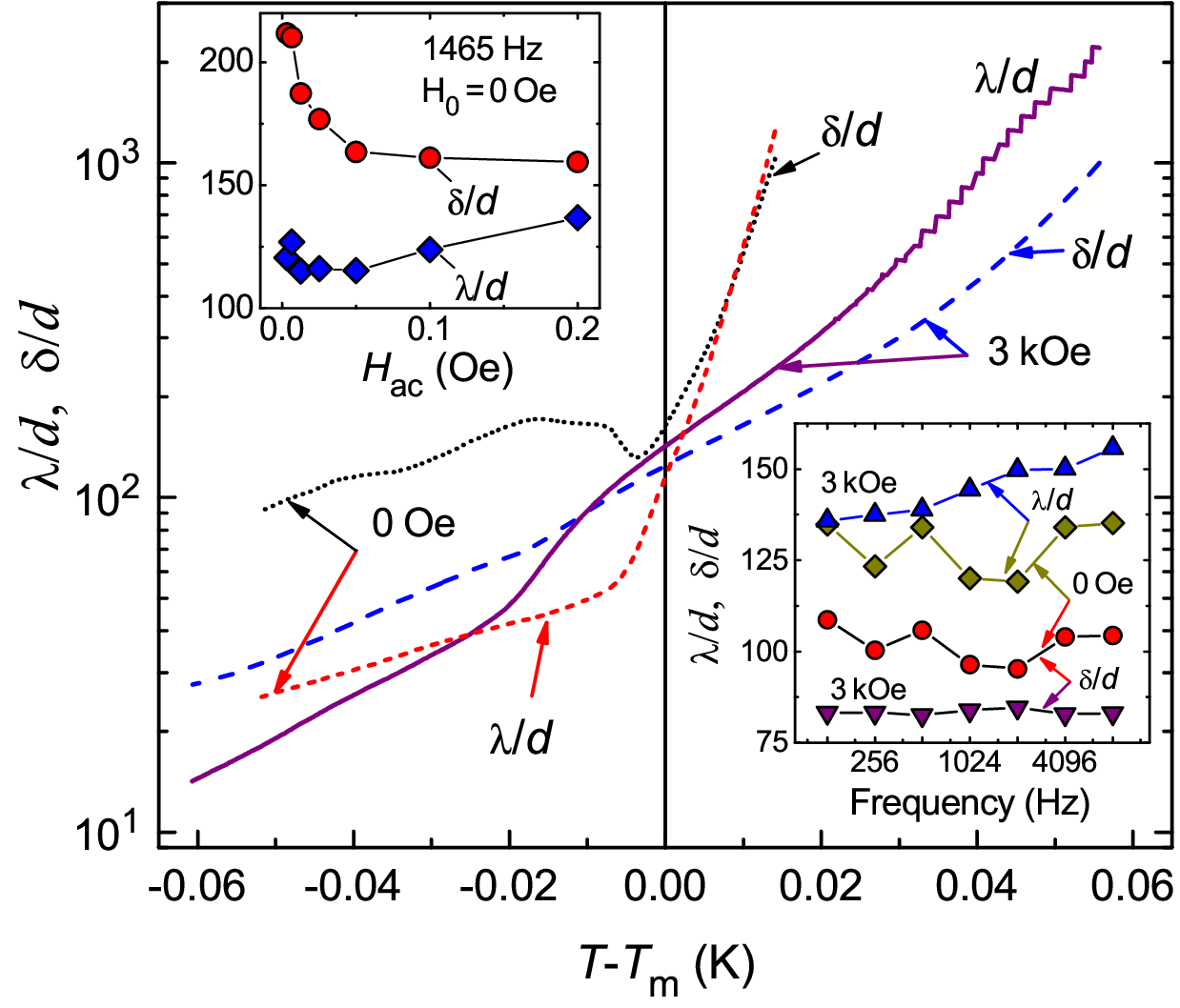}
    \caption{Temperature dependence of $\lambda/d$ and $\delta/d$ in zero and 3\,kOe dc magnetic field at a frequency of 1465\,Hz and $H_\mathrm{ac}= 0.05$\,Oe.
    Amplitude (upper inset) and frequency (lower inset) dependences of $\lambda/d$ and $\delta/d$ at $T=T_\mathrm{m}$.}
    \label{f5}
\end{figure}

Equation\,(\ref{Eq3}) permits to calculate the susceptibility $\chi_1$ of the slab. Inverting Eq.\,(\ref{Eq3}) one can find $\lambda$ and $\delta$ from the experimental data for $\chi_1$ at a given amplitude and frequency of the ac excitation. The obtained results are presented in Fig.\,\ref{f5}. For this plot, we used the experimental data obtained for sample GL at a frequency of 1465\,Hz and $H_\mathrm{ac} = 0.05$\,Oe at zero and 3\,kOe dc magnetic fields.

The amplitude dependences of $\lambda/d$ and $\delta/d$ at $T = T_\mathrm{m}$ are shown in the upper inset of Fig.\,\ref{f5}. The first thing to note is the large $\lambda/d$ and $\delta/d$ values just above $T_\mathrm{m}$. During the transition, $\delta/d$  decreases approximately by 10 times. That corresponds to an increase of the dissipative conductivity by 100 times, if one uses the standard expression for the normal skin-effect. Since the relaxation time of normal electrons is very short, $\omega \tau\ll1$, the frequency dispersion of the conductivity of such electrons is negligible at low frequencies, so that $\delta$ should be $\propto 1/\sqrt\omega$. However, there is no signatures of $\delta\propto 1/\sqrt\omega$ (lower inset in Fig.\,\ref{f5}). These data demonstrate the accuracy of the \emph{quasilinear} approach. For a \emph{linear} system its parameters do not depend on the amplitude of the external field. These results indicate that the observed $\chi_1^{\prime\prime}(\omega)$ is not directly related to the presence of the normal phase.

It should be noted that the results for $\lambda/d$ and $\delta/d$ given above were obtained in the linear approximation and they give only the estimated $\lambda/d$ and $\delta/d$. The large values of $\lambda/d$ and $\delta/d$ obtained within the framework of the two-fluid model, and especially the behavior of the dissipative part of the conductivity upon temperature and frequency variations, cannot explain the experimentally observed ac response, pointing to a different ac loss mechanism.

Another mechanism for the losses in a thin-walled cylinder can be associated with the penetration of an ac field into the slab. Therefore, it is desirable to analyze the stability of the superconducting state in a thin-walled cylinder, and the Ginzburg-Landau approach is well suited for this purpose.

\subsection{Ginzburg-Landau approach}
The stability of the superconducting state can be analyzed using the Ginzburg-Landau approach, taking into account that the sample thickness is much less than its length and width. For simplicity, we will also neglect end effects. The solution to this problem will be obtained similarly to the solution for surface superconducting states for an infinite plate in a parallel magnetic field, which was obtained more than 60 years ago\,\cite{Gen66boo,Sai69boo}. In what follows, compared to the approximation in Refs. \,\cite{Sai69boo}, the Ginzburg-Landau equations have not only a linear, but also a nonlinear term.

\begin{figure}[t!]
    \centering
    \includegraphics[width=0.85\linewidth]{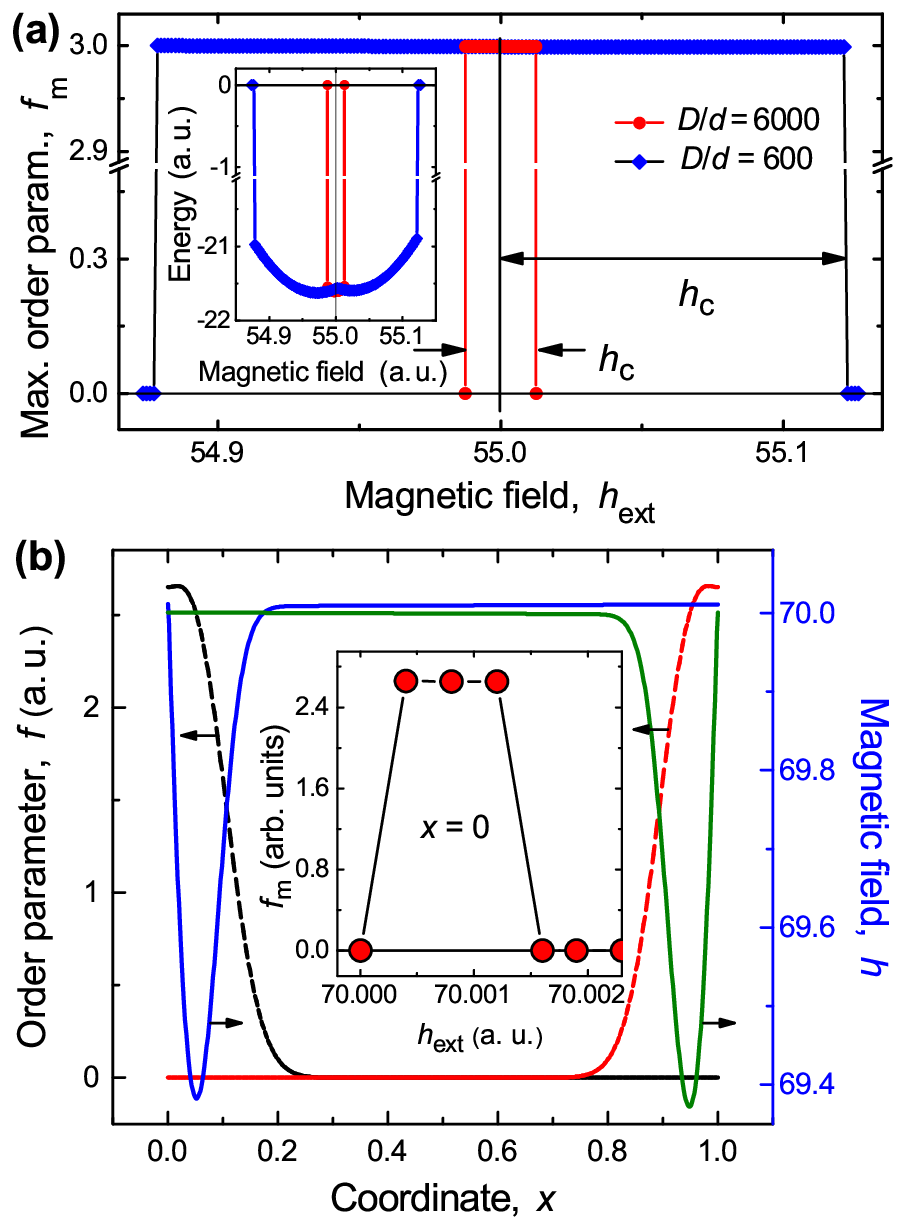}
    \caption{(a) Field dependence of the maximal superconducting order parameter $f_\mathrm{m}$ for two $D/d$ ratios at $h < h_\mathrm{c2}$. Inset: Field dependence of the energy of the system for the same $D/d$ ratios.
    (b) Order parameter $f$ and magnetic field $h$ in the regime $h > h_\mathrm{c2}$ as a function of the coordinate $x$.
    Inset: Field dependence of $f_\mathrm{m}$ at $x =0$.}
     \label{f6}
\end{figure}

The Ginzburg-Landau equations for the order parameter and the vector-potential in the film read
\begin{equation}
    \begin{array}{l}
    \label{Eq4}
     \displaystyle\frac{d^2f}{\kappa^2dx^2}-(\tilde{a}-k)^2f+ \displaystyle\frac{\xi_{0}^{2}}{\xi^2(T)}f-f^3=0, \\
     \\
     \displaystyle\frac{d^{2}\tilde{a}}{dx^{2}}-f^2(\tilde{a}-k) =0,
\end{array}
\end{equation}
with the order parameter $\psi=f(x)\psi_0\exp(iky)$, where $\psi_0$ is the order parameter and $\xi_0$ is the coherence length at arbitrary chosen temperature $T_0$, $\kappa = \lambda_0/\xi_0$  is the Ginzburg-Landau parameter, $\lambda_0 $  is London's penetration length at $T_0$ and $x=x/\lambda_{0}$ and $y=y/\lambda_0$ are the dimensionless coordinates. The dimensionless variables in Eq.\,(\ref{Eq4}) are the vector-potential $\tilde{a}=2e\xi_0A/\hbar c$, magnetic field $h=2e\xi_0\lambda_0H/\hbar c=2\pi \xi_0\lambda_0H/\phi_0$, $\phi_0$ is the flux quantum.
The chosen gauge is $\tilde{a}\mid_{x=0}=0$ in the center of the slab, since $x =0$ is the plane of symmetry of the problem, and the boundary conditions for the order parameter are $f$, $df/dx\mid_{ x=D}=0$ and $df/dx\mid_{x= D+d}=0$ in conjunction with the condition of continuity of the magnetic field on both sides of the films. The parameter $k$ determines the magnetic field inside the slab. In accordance with the Ginzburg-Landau equation, Eq.\,(\ref{Eq4}), $f$ will be nonzero only if $(\tilde{a}-k)^2\approx 0$. The internal magnetic field is uniform in the slab and we have $\tilde{a}_{\mid ^{ x=D}} = h_\mathrm{int}D\approx k$. When $k=h_\mathrm{ext}(D+d/2)$ the internal field is approximately equal to the external field, $h_\mathrm{int}\cong h_\mathrm{ext}$. The transition to the superconducting state corresponds to a change of the parameter $g=\xi_0^2/\xi^2(T)=(T_\mathrm{c}-T)/(T_\mathrm{c}-T_0)$ from zero to a certain final value. In the following we assume that the temperature $T_0$ was chosen from the condition $\lambda_0\equiv\lambda(T_0)=d$ at $\kappa=4$, $D/d=6250$ and $h_\mathrm{c2}=64$.
A numerical solution of Eqs. (\ref{Eq4}) reveals that the order parameter $f(x)$ in low magnetic fields weakly depends on $x$ and it has a maximum in the middle of the film. At the same time, from the solutions of Eqs.\,(\ref{Eq4}) follows that the superconducting state for a given magnetic field inside the slab (i.e., for a given $k$) can exist only in a narrow interval of external magnetic fields $|h_\mathrm{ext}-h_\mathrm{int}|<h_\mathrm{c}$. An increase of the external field outside this interval leads to an abrupt decrease of the order parameter to zero. The range of external magnetic fields, where the maximal order parameter $f_\mathrm{m}$ differs from zero, decreases rapidly with the ratio $D/d$. Figure\,\ref{f6}(a) shows the dependence of $f_\mathrm{m}$ on the external magnetic field $h_\mathrm{ext}< h_\mathrm{c2}$ for two ratios $D/d$. It can be seen from Fig.\,\ref{f6}(a) that $H_\mathrm{c}$ sharply decreases with increase of the ratio $D/d$. The inset in Fig.\,\ref{f6}(a) shows how the energy of the system depends on the magnetic field. The energy has two minima, which correspond to the SSS on opposite sides of the film.

Experiment reveals that the loss peak is located in high magnetic fields near $H_\mathrm{c3}$. Therefore, the spatial variation of the order parameter $f$ and the magnetic field $h$ in the film is of interest in external magnetic fields $h > h_\mathrm{c2}$. Figure\,\ref{f6}(b) shows how the order parameter $f$ and the magnetic field $h$ depend on the coordinate $x$ in the film for two very close external magnetic field values $h_\mathrm{ext}=70.00$ and $h_\mathrm{ext}=70.01$. The dashed lines correspond to the order parameter $f$ and the solid lines correspond to the magnetic field $h$. In Fig.\,\ref{f6}(b), the lines exhibiting strong nonlinearities near $x = 0$ correspond to $h_\mathrm{ext} = 70.00$ while those varying fast near $x = 1$ correspond to $h_\mathrm{ext} = 70.01$.  Note that for the selected film parameters, the field values are larger than $h_\mathrm{c2}=64$. Figure\,\ref{f6}(b) shows that the middle of the film is in the normal state with $f=0$ and $h=h_\mathrm{ext}$, and the superconducting state is located near the inner ($x=0$) and outer ($x=1$) surfaces of the film. The field dependence of the order parameter $f$ is shown in the inset of Fig.\,\ref{f6}(b). Here again, a very slight change of the external magnetic field strongly affects the state of the film.

This abrupt destruction of the superconducting state is analogous to the well-known current-induced breakdown of superconductivity in thin films\,\cite{Deg66boo}. However, the value of the critical current for a thin-walled cylinder is much smaller. The dynamics of the abrupt destruction of the superconducting state could be analysed in framework time-dependent Ginzburg-Landau model. But this problem is beyond the scope of our paper.

\subsection{Partial penetration of magnetic flux model}

The obtained solution of the Ginzburg-Landau equations shows that the critical field value for a thin-walled cylinder is very small. Therefore, $H_\mathrm{ac}$ of a small amplitude can destroy the superconducting state. The experiment reveals that the frequency dispersion of the ac response is negligible, see Fig.\,\ref{f4}(a), and therefore the quasistatic approach can be applied for the description of the ac response. In this case, application of the critical state model leads to the following scenario of the penetration of an ac field into a thin-walled cylinder. The ac component of the external magnetic field does not penetrate into the slab if $H_\mathrm{ac}<H_\mathrm{c}$. However, if $H_\mathrm{ac}\sin(\omega t)>H_\mathrm{c}$ at some time $t$, then the SSS becomes unstable, a transition to the normal state occurs, and the field penetrates into the slab. But, since $T<T_\mathrm{c}$, the normal state is unstable and the superconducting state is restored with a new value of $k$, which corresponds to an increase of the internal field. If full penetration occurs, the internal field becomes equal to the external one. Within the framework of the FPMF model, one can expect the appearance of jumps in the ac susceptibility as a function of the excitation amplitude, see Fig.\,\ref{f1}(b). These features have not been experimentally confirmed in the past. Neither they have been observed in our experiment, see Fig.\,\ref{f3}(d).

\begin{figure}[t!]
    \centering
       \includegraphics[width=0.9\linewidth]{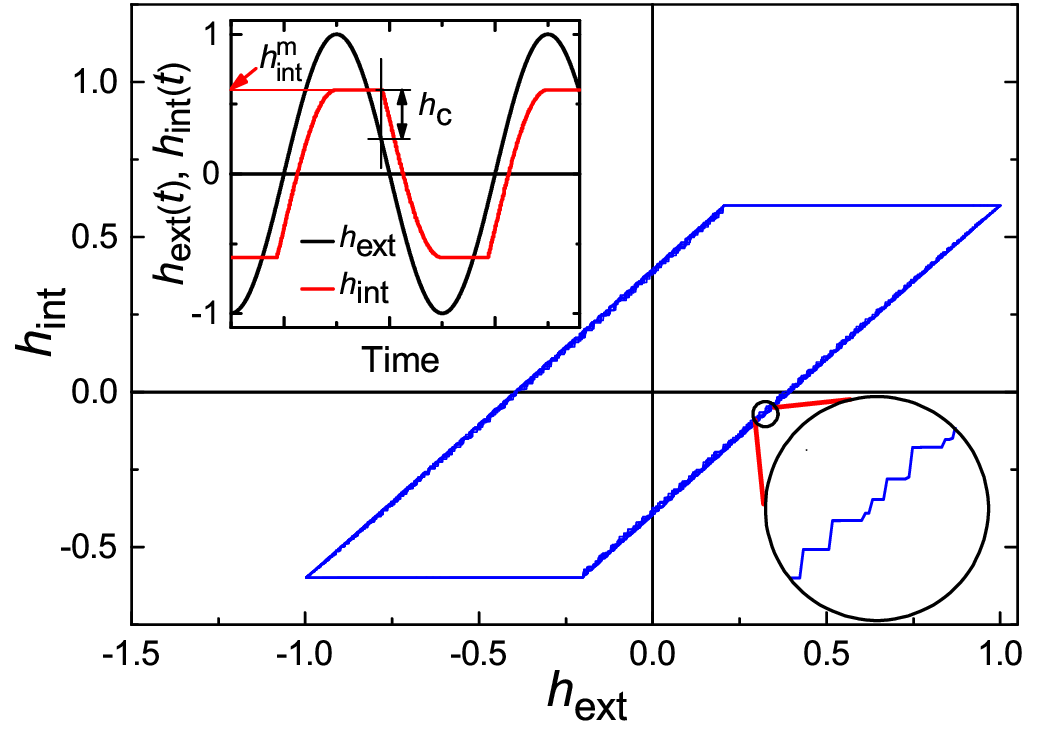}
    \caption{$h_\mathrm{int}(h_\mathrm{ext})$ during a slow course of 10 ac cycles. Zoom: tiny jumps of $h_\mathrm{int}(h_\mathrm{ext})$ in a part of a single ac cycle. Inset: Waveform of the ac magnetic fields $h_\mathrm{ext}(t)$ and $h_\mathrm{int}(t)$. $h_\mathrm{int}(t)$ is plotted after averaging over small jumps.}
     \label{f6new}
\end{figure}

At the same time, there is another possibility related to a \emph{partial penetration of magnetic flux} (PPMF) as follows. After restoration of superconductivity, the internal field does not reach the external field value, and the difference between them $|H_\mathrm{ext}-H_\mathrm{int}|/H_\mathrm{c}$ is a random variable value between $0$ and $1$. Within the framework of this model we obtain $H_\mathrm{int}$ as a function of the instant value of $H_\mathrm{ext}=H_\mathrm{ac}\sin(\omega t)$, in a form similar to that postulated in\,\cite{Rol67prv} for the ac response of bulk superconductors in the SSS. In the case of PPMF, the $H_\mathrm{int}(H_\mathrm{ext})$ loop during a slow passage of 10 ac cycles, looks as shown in Fig.\,\ref{f6new}.
\begin{figure*}
\centering
    \includegraphics[width=0.76\linewidth]{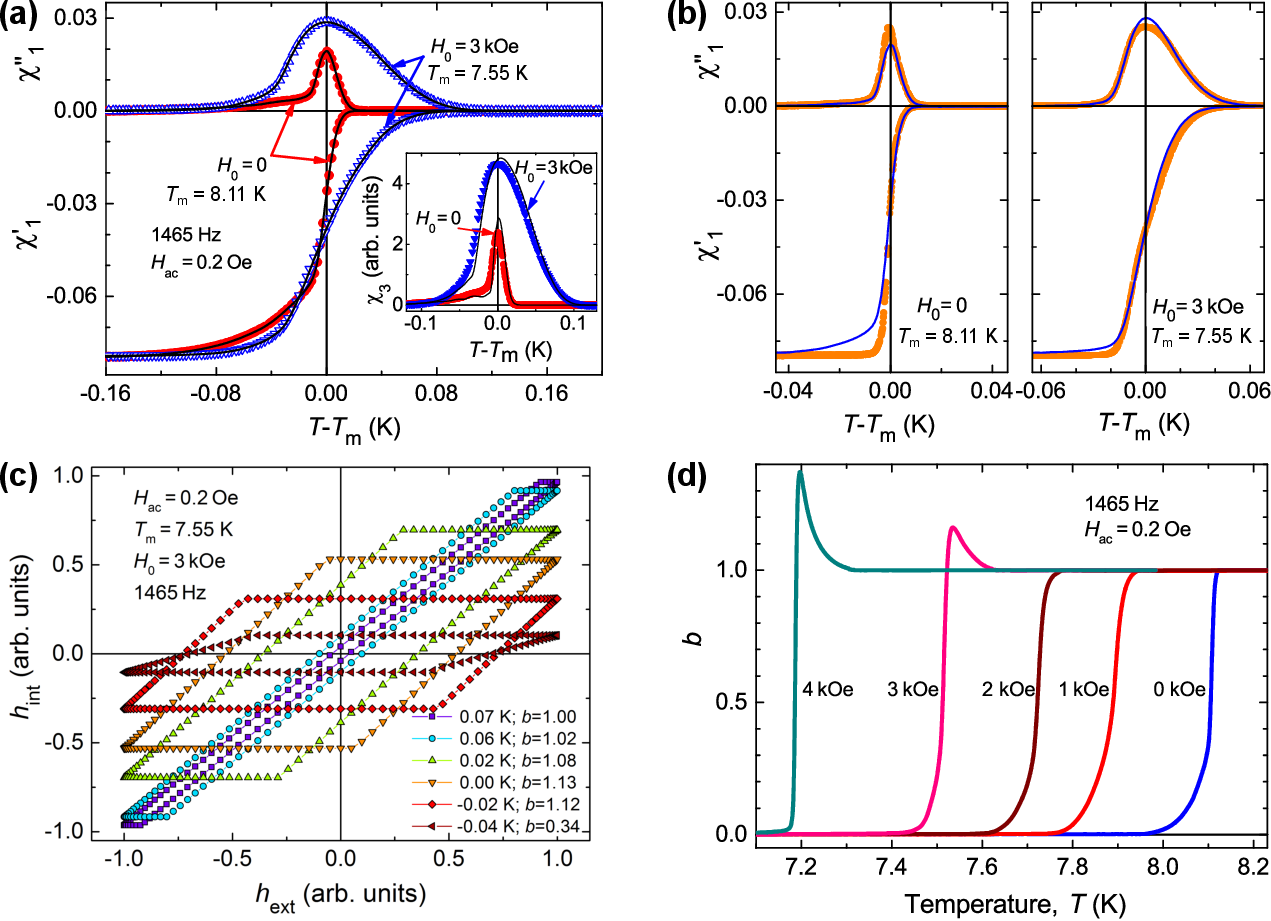}
    \caption{(a) $\chi_1^{\prime}(T)$ and $\chi_1^{\prime\prime}(T)$ (main panel) and $\chi_3(T)$ (inset) at zero and 3\,kOe dc magnetic fields. Symbols: experiment; lines: fits.
             (b) $\chi_1^{\prime}(T)$ and $\chi_1^{\prime\prime}(T)$ for $b=1$ at zero and 3\,kOe dc magnetic fields. Symbols: experiment; lines: fits.
             (c) Evolution of the $H_\mathrm{int}(H_\mathrm{ext})$ dependence with the temperature difference $(T-T_\mathrm{m})$, as indicated in the legend.
             (d) Temperature dependences of the parameter $b$ for a series of dc magnetic fields. In all panels, the measurements were carried out in the temperature-sweep mode. In all panels, the shown data are for sample GL.
    }
    \label{f7}
\end{figure*}
The enlarged area in Fig.\,\ref{f6new} shows the $H_\mathrm{int}(H_\mathrm{ext})$ dependence in a part of one ac cycle. The main difference between the RSF\,\cite{Rol67prv} and the PPMF model considered here is that in the RSF model (see Fig.\,\ref{f1}(a)) the field penetrates smoothly, whereas in the PPMF model small jumps are formed, see Fig.\,\ref{f6new}. It should be noted that a swept dc field can also penetrate into a thin-walled superconducting niobium cylinder in the form of giant jumps\,\cite{Tsi16pcs}. The presence of small jumps leads to the appearance of an excess noise at high frequencies, but this weakly affects the response at low frequencies, due to the fact that the ac response is measured with a lock-in amplifier which averages the signal over at least an hundred of ac cycles. After averaging over small jumps, the loop $H_\mathrm{int}=H_\mathrm{int}(H_\mathrm{ext})$ in Fig.\,\ref{f6new} can be characterized by two free parameters: (1) $a=H_\mathrm{int}^\mathrm{m}/H_\mathrm{ac}$, where $H_\mathrm{int}^\mathrm{m}$ is the maximal value of $H_\mathrm{int}$ (see the inset in Fig. \ref{f6new}) and (2) the average slope of the inclined part of the $H_\mathrm{int}(H_\mathrm{ext})$ loop, $b=dH_\mathrm{int}/dH_\mathrm{ext}$. For the RSF model\,\cite{Rol67prv} and the model suggested here the parameter $b = 1$. It should be noted that random changes in $|H_\mathrm{ext}-H_\mathrm{int}|/H_\mathrm{c}$ are necessary since jumps will appear in the amplitude dependence of the ac susceptibility for any constant value of this ratio less than $1$.

\section{Discussion}
The relation between the internal and external magnetic fields obtained for thin-wall slabs within the framework of the PPMF model is employed for fitting the experimentally measured dependences of the susceptibility on temperature, dc magnetic field and amplitude of excitation. To accomplish this, one proceeds as follows. First, after averaging the minor magnetization loop over small jumps, one obtains the waveform of $H_\mathrm{int}(t)$ as shown in the inset of Fig.\,\ref{f6new}. Next, by performing the Fourier transform, the in-phase and out-of-phase components of the ac susceptibility can be obtained as a function of the parameters $a$ and $b$. Finally, by varying $a$ and $b$ as fitting parameters, it is possible to fit the experimental data for $\chi_1 (T)$ near the superconducting transition at any temperature point. The result of the fitting procedure is shown in
Fig.\,\ref{f7}(a). One can see that the fits are in excellent agreement with the experimental data. The temperature dependence for the parameter $b$ is shown in Fig.\,\ref{f7}(d).

The parameters $a$ and $b$ obtained from fitting of $\chi_1$ can be used for the calculation of $\chi_3$. The criterion for this calculation is the equality of the areas under the experimental and calculated curves. Figure\,\ref{f7}(a) shows the fits for the temperature dependence of the first-harmonic ac susceptibility. It turns out that the third harmonics of $H_\mathrm{int}(H_\mathrm{ext},a,b)$ with $a$ and $b$ obtained from the above mentioned procedure does not fit very well to the measured $\chi_3(T)$ dependences. Nevertheless, the general behavior of $\chi_3(T)$ is reproduced quite well. It should be noted that the  $H_\mathrm{int}(H_\mathrm{ext})$ dependence with the parameter $b = 1$ can also be used for fitting of the experimental data. As an example, in Fig.\,\ref{f7}(b) we show the best fits with only one free parameter $a$. These fits have a larger discrepancy around the superconducting transition.

The performed calculations allow us to plot the  $H_\mathrm{int}(H_\mathrm{ext})$ dependence at any temperature in the temperature range where $\chi_1^{\prime\prime}$ is nonzero. The shape of the $H_\mathrm{int}(H_\mathrm{ext})$ dependence changes with temperature, as shown in Fig.\,\ref{f7}(c). It can be seen from Fig.\,\ref{f7}(c) that the parameters $a$ and $b$ change with temperature. It is natural that in the normal state, the parameters $a =1$ and $b =1$ since $H_\mathrm{int} = H_\mathrm{ext}$. Deeply in the superconducting state, the parameters $a$ and $b$ turn to zero since $H_\mathrm{int}$ = const. Figure\,\ref{f7}(d) shows the temperature dependence of the parameter $b$ for a series of dc magnetic fields values. In addition, when $H_0>2$\,kOe, the parameter $b$ near $T_\mathrm{m}$ exceeds 1. It turns out that the peak magnitude of the parameter $b$ increases with increasing excitation amplitude.

\begin{figure}
    \centering
    \includegraphics[width=1\linewidth]{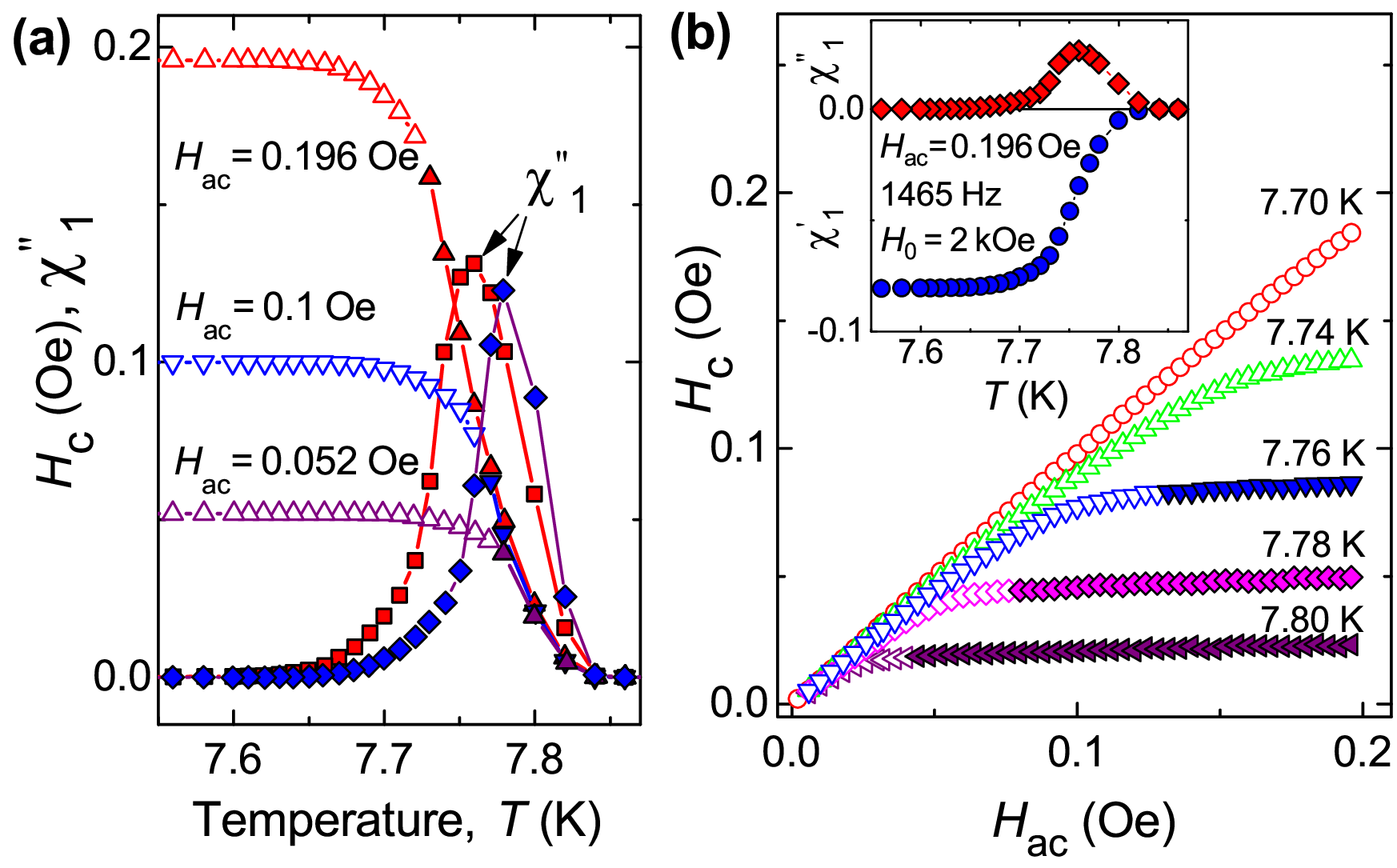}
    \caption{(a) Temperature dependence of the critical magnetic field $H_\mathrm{c}$ for three excitation amplitudes and $5\times\chi_1^{\prime\prime}(T)$ at $H_\mathrm{ac} = 0.196$\,Oe (red squares) and 0.1\,Oe (blue diamonds). Filled symbols in the $H_\mathrm{c}(T)$ curves correspond to the range of reliable determination of $H_\mathrm{c}$.
    (b) Amplitude dependence of $H_\mathrm{c}$ at temperatures near the absorption peak, $T_\mathrm{m} = 7.76$ K at $H_\mathrm{ac} = 0.196$ Oe.
    Filled symbols corresponds to the range of reliable determination of $H_\mathrm{c}$.
    Inset: Temperature dependences of $\chi_1^{\prime}$ and $\chi_1^{\prime\prime}$ at $H_0 = 2$\,kOe and $H_\mathrm{ac} = 0.196$\,Oe.
    In both panels, the measurements were carried out in the point-by-point mode.}
    \label{f8}
\end{figure}

Finally, it is possible to plot the critical field $H_\mathrm{c} = h_\mathrm{c} H_\mathrm{ac}$ as a function of the external parameters such as temperature and/or excitation amplitude. Here, $H_\mathrm{ac}$ is the amplitude of the applied ac magnetic field in oersted and $h_\mathrm{c}\equiv(1-a)$ is defined within the framework of the PPMF model. Figure\,\ref{f8}(a) shows the temperature dependences of the critical field $H_\mathrm{c}(T)$ and $\chi_1^{\prime\prime}(T)$ at a magnetic field of 2\,kOe for ac amplitudes of 0.196, 0.1, and 0.052\,Oe. The imaginary part of the ac susceptibility was measured at $H_\mathrm{ac} = 0.196$\,Oe and $0.1$\,Oe. The measurements were carried out in the point-by-point mode with decreasing temperature. This figure shows that $H_\mathrm{c}$ increases with decreasing temperature, as expected, and it saturates at low temperatures. The beginning of empty symbols with decreasing temperature corresponds to the beginning of $H_\mathrm{c}$ saturation. Note that calculation of $H_\mathrm{c}$ is meaningful when $H_\mathrm{c}<H_\mathrm{ac}$ whereas at sufficiently low temperatures $H_\mathrm{c}$ becomes greater than $H_\mathrm{ac}$. It should be noted that $H_\mathrm{c}$ was calculated using the parameter $a$ obtained from fitting the experimental data with both $a$ and $b$ varied as two fitting parameters. The temperature dependence of the parameter $a\equiv 1 - h_\mathrm{c}$ can be deduced from the $H_\mathrm{c}(T)$ plot in Fig.\,\ref{f8}(a).

We emphasize that calculation of the parameters of the $H_\mathrm{int} (H_\mathrm{ext})$ loops, and then of $H_\mathrm{c}$, is only possible if both $\chi_1^{\prime}$ and $\chi_1^{\prime\prime}$ are nonzero. However, at low temperatures $\chi_1^{\prime\prime}=0$. At high temperatures, $H_\mathrm{c}(T)$ is a quadratic rather than a linear function of temperature, as one might expect. In Fig.\,\ref{f8}(b) we show the amplitude dependences of $H_\mathrm{c}(H_\mathrm{ac})$ at constant temperatures near the absorption peak. It is seen that at $T\geq T_\mathrm{m}$, $H_\mathrm{c}$ reaches a plateau, i.e. ceases to depend on $H_\mathrm{ac}$, and as the temperature decreases, the plateau is reached at higher amplitudes. Naturally, $H_\mathrm{c}$ is a parameter of the system and should not depend on the excitation amplitude. Hence, one can conclude that the calculations of $H_\mathrm{c}$ from the experimental data are reliable only at $T\geq T_\mathrm{m}$ in the plateau region.

In all, the $H_\mathrm{int} (H_\mathrm{ext})$ loops calculated within the framework of the PPMF model are in good agreement with the experimental data in the entire temperature range where both $\chi_1^{\prime}$ and $ \chi_1^{\prime\prime}$ are nonzero, see Fig.\,\ref{f7}(a). However, it is impossible to conclude that this model fully describes the experiment for any dc fields and excitation amplitudes, since there is a region of fields and temperatures where the parameter $b\neq 1$. Note that $b>1$ means that the change in the internal field is larger than that of the external one. However, for $T>T_\mathrm{m}$, the parameters $a$ and $b$ of the $H_\mathrm{int} (H_\mathrm{ext})$ dependences are reasonable and the model of the $H_\mathrm{int} (H_\mathrm{ext})$ loops is fully applicable.

The performed analysis can also be discussed in the context of applicability of other models. For instance, an alternative ac loss mechanism could be, in principle, associated with cylinder end effects. This assumption implies that the concentration of magnetic force lines close to the top and bottom cylinder rims might lead to the penetration of tilted vortex segments. An increase in the field will lead to the movement of the vortex sections to the center of the cylinder, where the vortices from top and antivortices from bottom cylinder will annihilate. This leads to flux penetration through the wall and losses. Let's assume that this scenario works in our case. It should be noted that the time scale in our experiment is determined by the period of the ac field. Thus, within half a period, in the frame of this model, the vortex should reach the center of the sample, and its velocity at a frequency of, say, 1 kHz should be more than 20 m/s. However, the calculated vortex velocity with an excitation amplitude of about $10^{-2}$ Oe and an average resistivity of the niobium film in the normal state of $14 \mu\Omega$cm and with a film thickness of 120 nm can reach $10\div 20$ cm/s. These values are the upper limit for the vortex velocity under conditions of our experiment. Consequently, the vortex and antivortex moving towards the center of the sample will never meet each other during half the ac period. The above estimates were made for the case when the vortex axis is perpendicular to the film surface. In addition, it is necessary to take into account that the magnetic force lines in the center of the sample are parallel to its axis. An assessment of the penetration of a parallel dc magnetic field into a thin-walled cylinder shows that at a distance from the cylinder rim exceeding $10$\%
its radius, the normal component of the magnetic field becomes negligible small\,\cite{Tsi19pcs}. The length of our samples significantly exceeds the radius of the cylinder.
It means, that tilted vortex and antivortex segments do not exist in the middle of the sample. From the above it follows that the model based on the movement of the vortex and antivortex from the edges of the sample is unlikely.

 It is clear that the most realistic model should be related to the experimental result that the loss peak is located near the $H_{c3}(T)$ line of a thin-walled cylinder. Consequently, such a model can be based on the solution of the problem of the nucleation of a superconducting phase in a plate in a parallel magnetic field. The solution to such a model was obtained long time ago in the linear Ginzburg-Landau approximation\,\cite{Sai69boo}. Subsequently, this solution was successfully applied to explain the reentrance of superconductivity in strong magnetic fields by stopping the movement of vortices due to the appearance of a superconducting surface sheath\,\cite{Cor13nac}. Accordingly, it can be assumed that the observed ac losses in our experiment can be explained by a further generalization of the solution obtained in Refs.\,\cite{Sai69boo,Cor13nac}. However, in the case of an ac experiment, stopping of the vortex motion near $H_\mathrm{c3}$ could lead to a splitting of the field or temperature dependences of $\chi_1 ^{\prime\prime}$. We note that such a splitting has not been observed experimentally either in this or our previous works\,\cite{Tsi14prb,Tsi16pcm}. Nevertheless, we believe that a new, more accurate model that could resolve the issues encountered in the model developed here could be based on the models described in Refs.\,\cite{Sai69boo} and \,\cite{Cor13nac}.  But such a solution does not yet exist.

\section{Conclusion}
To summarize, we have investigated experimentally the ac response of thin-walled hollow superconducting cylinders in the course of their transition to the superconducting state when both ac and dc magnetic fields are applied parallel to the cylinder axis. We demonstrate that the ac susceptibility $\chi_{1}$ of the thin-walled cylinders is a smooth function of the amplitude of the applied ac magnetic field. However, this experimental finding is in contrast to the prediction of the critical state FPMF model which implies full penetration of magnetic flux. Namely, the FPMF model asserts that $\chi_{1}(H_\mathrm{ac})$ should exhibit jumps because of the periodic destruction and restoration of the surface superconducting state in the cylinder wall.

Our theoretical analysis within the framework of the two-fluid model has revealed that the presence of normal component of the conductivity cannot explain the experimentally observed behavior of ac losses, suggesting the penetration of the ac field into the cylinder hollow as another possible mechanism. The modeling based on the Ginzburg-Landau equation has shown that indeed, at magnetic fields exceeding the second critical field, the superconducting state is very sensitive to even small variations of the external ac field, which can penetrate into the cylinder hollow.

For settling the inconsistency between the prediction of jumps in $\chi(H_\mathrm{ac})$ in the FPMF model and the smooth $\chi(H_\mathrm{ac})$ observed experimentally, we have proposed a phenomenological model of partial penetration of magnetic flux, PPMF.  This model implies that after a restoration of the superconducting state, the magnetic fields inside and outside the cylinder are not equal, and the value of the penetrated flux is random for each penetration. In this case $\chi_{1}(H_\mathrm{ac})$ is smooth and the model of minor magnetization loops $H_\mathrm{int}(H_\mathrm{ext})$ fits very well the experimental data for the temperature dependence of the first-harmonic $\chi_1$, at any dc magnetic fields and excitation amplitudes. However, at temperatures below $T_\mathrm{m}$ the deduced physical parameters should be treated with care. It should be noted that the PPMF model shows how the critical state model with full penetration of magnetic flux could be transformed into the Rollins-Silcox-Fink model.

\vspace{4mm}
\section*{Acknowledgements}
We thank Yu. A. Genenko and G. I. Leviev for helpful discussions. This article is based upon work from COST Action CA21144 (SuperQuMap), supported by COST (European Cooperation in Science and Technology). Financial support of the grant agency VEGA 2/0140/22 and APVV 19-0303 is kindly appreciated. OVD acknowledges financial support by the Austrian Science Fund (FWF) via Grant Nos. I 4889 (CurviMag) and I 6079 (FluMag). Further, work of OD was funded by the Deutsche Forschungsgemeinschaft (DFG, German Research Foundation) under Germany's Excellence Strategy - EXC-2123 QuantumFrontiers - 390837967.


%

\end{document}